\documentstyle[12pt,twoside]{article}
\input{epsf}
\begin{document}
\setlength{\baselineskip}{12pt}
\hoffset 0.65cm
\voffset 0.3cm
\bibliographystyle{normal}

\vspace{48pt}
\noindent
{\bf Flexoelectric Instability in Nematic Liquid Crystal
between Coaxial Cylinders}

\vspace{48pt}
\noindent
I.~V. KOTOV$^a$, M.~V. KHAZIMULLIN$^a$ and A.~P. KREKHOV$^{a,b}$

\noindent
$^a$Institute of Molecule and Crystal Physics, Russian Academy of 
Sciences, 450025 Ufa, Russia;
$^b$Physikalisches Institut, Universit\"at Bayreuth,
D-95440 Bayreuth, Germany

\vspace{36pt}
\noindent
The stability of the equilibrium configurations of a nematic liquid 
crystal confined 
between two coaxial cylinders is analysed when a radial electric 
field is applied and the flexoelectric effect is taken into account.
The threshold for perturbations 
depending only on the radius $r$ in the cylindrical coordinate system
and strong boundary conditions is studied.
A new type of orientational transition caused by pure flexoelectric
effect is found.

\vspace{24pt}
\noindent
\underline{Keywords:}~nematic liquid crystal, flexoelectric
instability

\vspace{36pt}
\baselineskip=1.25 \baselineskip

\noindent
{\bf INTRODUCTION}
\vspace{12pt}

\noindent
The uniform orientation of nematic liquid crystal (NLC) is caused by
the orienting action of appropriately treated confining plates,
which define a fixed orientation of the NLC at the boundary.
When an electric or magnetic field is applied to such NLC sample
in an appropriate direction, a director reorientation via a
Fr\'eedericksz transition may result, which has been studied extensively
in case of a plane layer geometry.

For a NLC confined between two coaxial cylinders
the orientational transition may occur even in the absence of external 
fields depending on the ratio of the radii of the inner cylinder to that
of the outer one \cite{CK72,WCK73}.
The Fr\'eedericksz transition in this geometry was analysed for a
magnetic field directed radially outwards from the common axis of the 
cylinders \cite{Leslie70} and tangentially to circles in the plane
perpendicular to the cylinders' axis \cite{Kini77,Tsuru90} as well as
for a radial electric field.
Here a voltage was applied between the
two coaxial cylinders \cite{AOY79,WH93}.
In this geometry the 
critical voltage for the Fr\'eedericksz transition depends on the ratio between
the cylinders radii.

In this paper we analyse the influence of the flexoelectric effect
on the orientational transitions 
in a NLC confined between two coaxial cylinders under applied d.c.
radial electric field.
The flexoelectricity describes the linear coupling between an applied
electric field and gradients in the director field \cite{Meyer69}.
In contrast to the case of a plane nematic layer, in cylindrical geometry
under applied electric field one has a contribution from the
flexoelectric effect in the bulk torque acting on the director.
Here we calculate the critical voltage for the orientational
transition where the director distribution 
${\bf \hat n}=(n_r, n_{\phi}, n_z)$ depends only on the radius $r$
in the cylindrical coordinate system $(r,\phi,z)$ with $z$ axis along
the axis of cylinders.
Strong boundary conditions are imposed.
Two typical initial director distributions [homeotropic orientation
${\bf \hat n}=(1, 0, 0)$ and planar one ${\bf \hat n}=(0, 1, 0)$]
are considered.
It is found that for planar orientation a new type of orientational
transition caused purely by the flexoelectric effect takes place.

\vspace{24pt}
\noindent
{\bf BASIC EQUATIONS}
\vspace{12pt}

\noindent
Let us consider the nematic enclosed between two infinite coaxial
cylindrical electrodes whose inner and outer radii are $r_1$ and 
$r_2$, respectively.
The director distribution ${\bf \hat n} = {\bf \hat n}(r)$ depends
only on $r$.
The voltage applied between the electrodes provides the radially
directed electric field ${\bf E} = (E,0,0)$.
The equilibrium director configuration can be found from the extremum
condition of the total free energy (per unit length of cylinders)
\begin{equation}
\label{te}
{\cal F} = 2 \pi \int_{r_1}^{r_2} (F_d+F_{el}+F_{fl}) r d r \;.
\end{equation}
Here $F_d$ is the elastic energy density
\begin{equation}
F_d = \frac{1}{2}
\Bigl[ K_{1}\left(\nabla\cdot{\bf \hat n}\right)^2+
K_{2}\left({\bf \hat n}\cdot \left(\nabla \times{\bf \hat n}\right) \right)^2+
K_{3}\left({\bf \hat n}\times\left(\nabla \times {\bf \hat n}\right) \right)^2 \Bigr] \;.
\end{equation}
The electric field contribution is given by
\begin{equation}
F_{el}=-\frac{1}{2}\epsilon_0\epsilon_{a}\left({\bf \hat n}\cdot{\bf E}\right)^2 \;,
\end{equation}
where $\epsilon_a=\epsilon_\parallel-\epsilon_\perp$ is the dielectric
anisotropy and the radial component of the electric field
$E = U/[r\ln(r_{2}/r_{1})]$ with $U=U(r_1)-U(r_2)$ the voltage 
drop across the cylinders.
The contribution of flexoelectricity
\begin{equation}
F_{fl}=-\left(\bf E\cdot\bf P\right) \; , \;\;
{\bf P} = e_{11}{\bf \hat n} \left(\nabla \cdot {\bf \hat n}\right) -
e_{33}\left({\bf \hat n}\times\left(\nabla \times {\bf \hat n}\right)\right)
\end{equation}
is connected with flexo-polarization ${\bf P}$
induced by the splay and bend deformations of the director field and
$e_{11}$, $e_{33}$ are the flexoelectric coefficients.

Minimisation of (\ref{te}) with the normalisation ${\bf \hat n}^2=1$
gives the torque balance equation
\begin{equation}
\label{torqbal}
\left[ {\bf \hat n} \times {\bf h} \right] = 0 \; , \;\; 
h_i = \frac{d}{d r} \left( \frac{\partial f}{\partial n_{i,r}} \right) 
- \frac{\partial f}{\partial n_i} \; , \;\; i=r, \phi, z
\end{equation} 
where $f=(F_d+F_{el}+F_{fl})r$.
One can easily verify that ${\bf \hat n}_0=(1,0,0)$ and ${\bf \hat n}_0=(0,1,0)$
are solutions of equation (\ref{torqbal}) for homeotropic and 
planar boundary conditions, respectively.

\vspace{24pt}
\noindent
{\bf STABILITY ANALYSIS}
\vspace{12pt}

\noindent
Let us now examine the stability of these configurations.

\noindent
{\bf I)} Homeotropic anchoring with boundary conditions
\begin{equation}
{\bf \hat n}(r=r_1) = {\bf \hat n}(r=r_2) = (1,0,0) \;.
\end{equation} 
Small deviations from the exact solution ${\bf \hat n}_0=(1,0,0)$ of
(\ref{torqbal}) can be written as
\begin{equation}
{\bf \hat n} = {\bf \hat n}_0 + \delta{\bf \hat n} = 
\left[ 1-\frac12 (\delta n_{\phi}^2 + \delta n_{z}^2); \delta n_{\phi}; \delta n_{z}  \right]
\end{equation} 
up to second order in the perturbations $\delta n_{\phi}$, $\delta n_{z}$ which depend 
only on $r$.
Then for the change of total free energy one obtain up to second order in
$\delta {\bf \hat n}$
\begin{eqnarray}
&& \Delta {\cal F}_{hom} \equiv 
{\cal F}({\bf \hat n}_0) - {\cal F}({\bf \hat n}_0+\delta {\bf \hat n}) = 
\nonumber\\
&& = 2 \pi \int_{r_1}^{r_2} 
\Bigl\{ 
\frac12 K_1 [-\frac{1}{r^2}(\delta n_{\phi}^2 + \delta n_{z}^2)] +
\frac12 K_3 [\delta n_{\phi}'^2 + \delta n_{z}'^2 + \frac{1}{r^2}\delta n_{\phi}^2] + 
\nonumber\\
&& \qquad \qquad \,\,\, 
\frac12 \epsilon_0\epsilon_a {\tilde U}^2 \frac{1}{r^2} (\delta n_{\phi}^2 + \delta n_{z}^2) + 
\nonumber\\
&& \qquad \qquad \,\,\, 
{\tilde U} e_{11} \frac{1}{r^2} (\delta n_{\phi}^2 + \delta n_{z}^2) + 
{\tilde U} e_{33} \frac{1}{r^2} \delta n_{\phi}^2 
\Bigr\} r d r \; ,
\end{eqnarray} 
where the boundary conditions $\delta {\bf \hat n}(r=r_1)=\delta {\bf \hat n}(r=r_2)=0$ 
were taken into account, ${\tilde U}=U/\ln(r_2/r_1)$ and $f' \equiv d f/d r$.

The necessary and sufficient condition for $\Delta {\cal F}_{hom}$ to be
positive definite is that the Euler-Lagrange equations
\begin{eqnarray}
\label{dnphi}
&& K_3 \frac{d^2}{d \alpha^2}\delta n_{\phi} +
[K_1 - K_3 - \epsilon_0\epsilon_a {\tilde U}^2 - 2{\tilde U}(e_{11}+e_{33})] \delta n_{\phi} = 0 \; , \;\; \\
\label{dnz_h}
&& K_3 \frac{d^2}{d \alpha^2}\delta n_{z} + 
[K_1 - \epsilon_0\epsilon_a {\tilde U}^2 - 2{\tilde U}e_{11}] \delta n_{z} = 0 \; ,
\end{eqnarray}
have nonzero solutions $\delta n_{\phi}, \delta n_{z} \ne 0$ in the range 
$0<\alpha<\ln(r_2/r_1)$ 
where we introduced the new variable $\alpha=\ln(r/r_1)$.
Since equations (\ref{dnphi}), (\ref{dnz_h}) are uncoupled one gets two independent conditions
for the instability of initial homeotropic orientation with respect to $\delta n_{\phi}$ and
$\delta n_{z}$ perturbations.
Solving (\ref{dnphi}), (\ref{dnz_h}) one obtains for the critical 
voltage $U_c$ for the two types of orientational transitions
in cylindrical geometry 
\begin{eqnarray}
\label{Uchdnphi}
\delta n_{\phi}\;: &&
-{\rm sign}(\epsilon_a)\left(\frac{U_c}{U_{F_3}}\right)^2 - 
\left(\frac{U_c}{U_{F_3}}\right) \frac{2(e_{11}+e_{33})\ln(r_2/r_1)}{\pi \sqrt{K_3\epsilon_0|\epsilon_a|}} = 
\nonumber\\
&&  \qquad \;\;\;\;\,\, = 1 + \frac{K_3-K_1}{K_3}\left(\frac{\ln(r_2/r_1)}{\pi}\right)^2 \; ,
\\
\label{Uchdnz}
\delta n_{z}\;: &&
-{\rm sign}(\epsilon_a)\left(\frac{U_c}{U_{F_3}}\right)^2 - 
\left(\frac{U_c}{U_{F_3}}\right) \frac{2e_{11}\ln(r_2/r_1)}{\pi \sqrt{K_3\epsilon_0|\epsilon_a|}} = 
\nonumber\\
&&  \qquad \;\;\;\;\,\,  = 1 - \frac{K_1}{K_3}\left(\frac{\ln(r_2/r_1)}{\pi}\right)^2 \; ,
\end{eqnarray}
where $U_{F_3}=\pi\sqrt{K_3/\epsilon_0|\epsilon_a|}$ is the threshold voltage for the bend 
Fr\'ee\-de\-ricksz transition in case of a plane NLC layer.
\begin{figure} 
\begin{center} 
\vspace*{-0.7cm} 
\hspace*{0cm} 
\epsfxsize=7.8cm 
\epsfbox{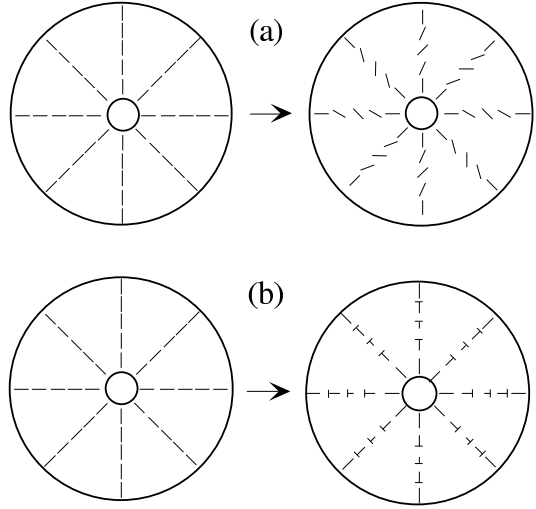} 
\end{center} 
\vspace*{-1.0cm}
\caption{Orientational transition for $\delta n_{\phi}$ (a) 
and $\delta n_{z}$ (b).}
\vspace*{0.1cm}
\label{hom} 
\end{figure} 
In Figure~\ref{hom} the two types of transitions correspond to 
$\delta n_{\phi}$ and $\delta n_{z}$
perturbations are shown schematically.
In the absence of electric field ($U=0$) there is a possibility of 
orientational transition for a radius ratio above some
critical value.
The transition corresponding to $\delta n_{z}$ perturbation [Fig.~1(b)] 
comes first and occurs at
$\ln(r_2/r_1)=\pi\sqrt{K_3/K_1}$.

\begin{figure}
\begin{center}
\vspace*{-0.9cm}
\hspace*{-0.5cm}
\epsfysize=4.9cm
\epsfbox{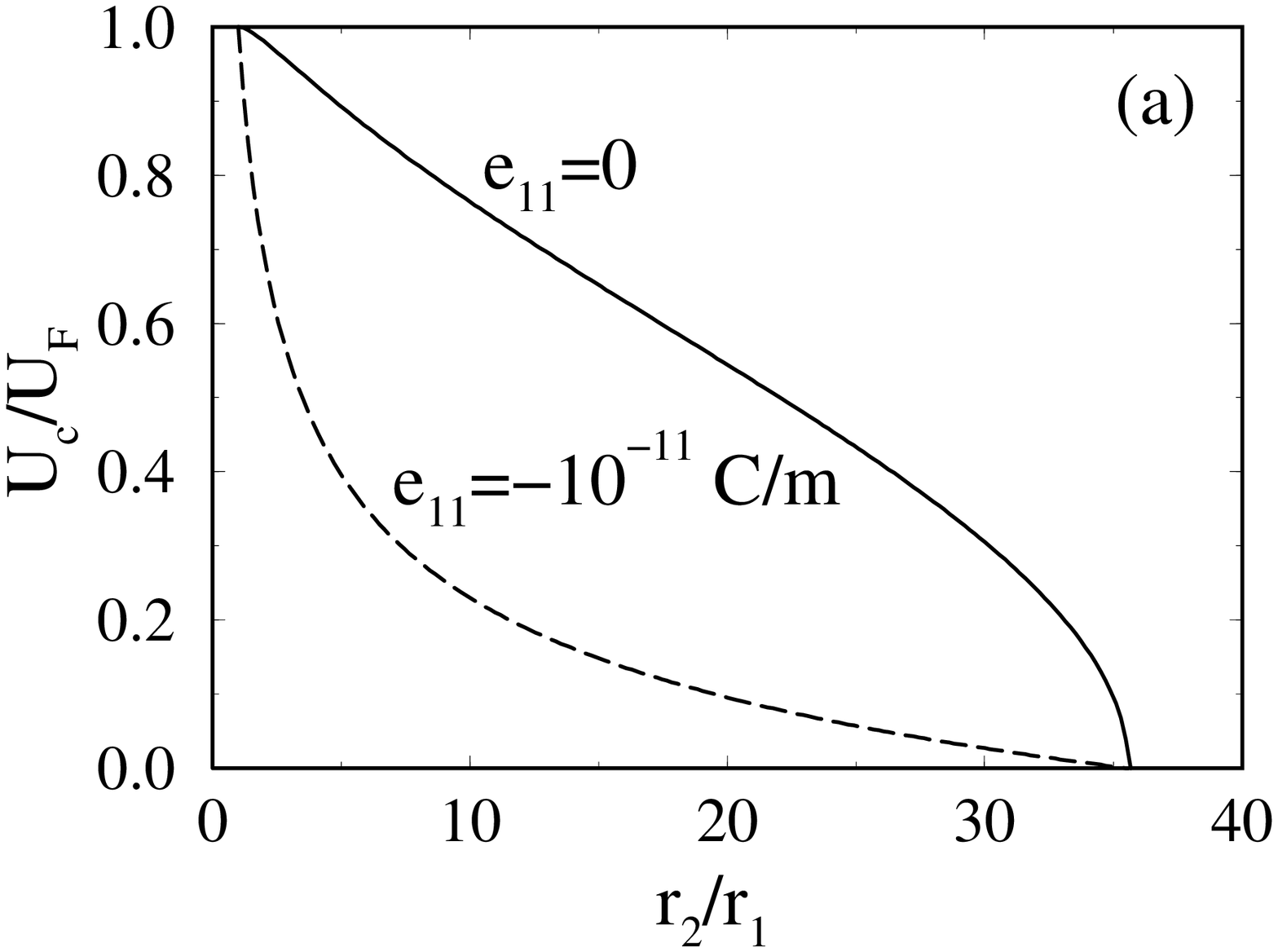}
\epsfysize=4.9cm
\epsfbox{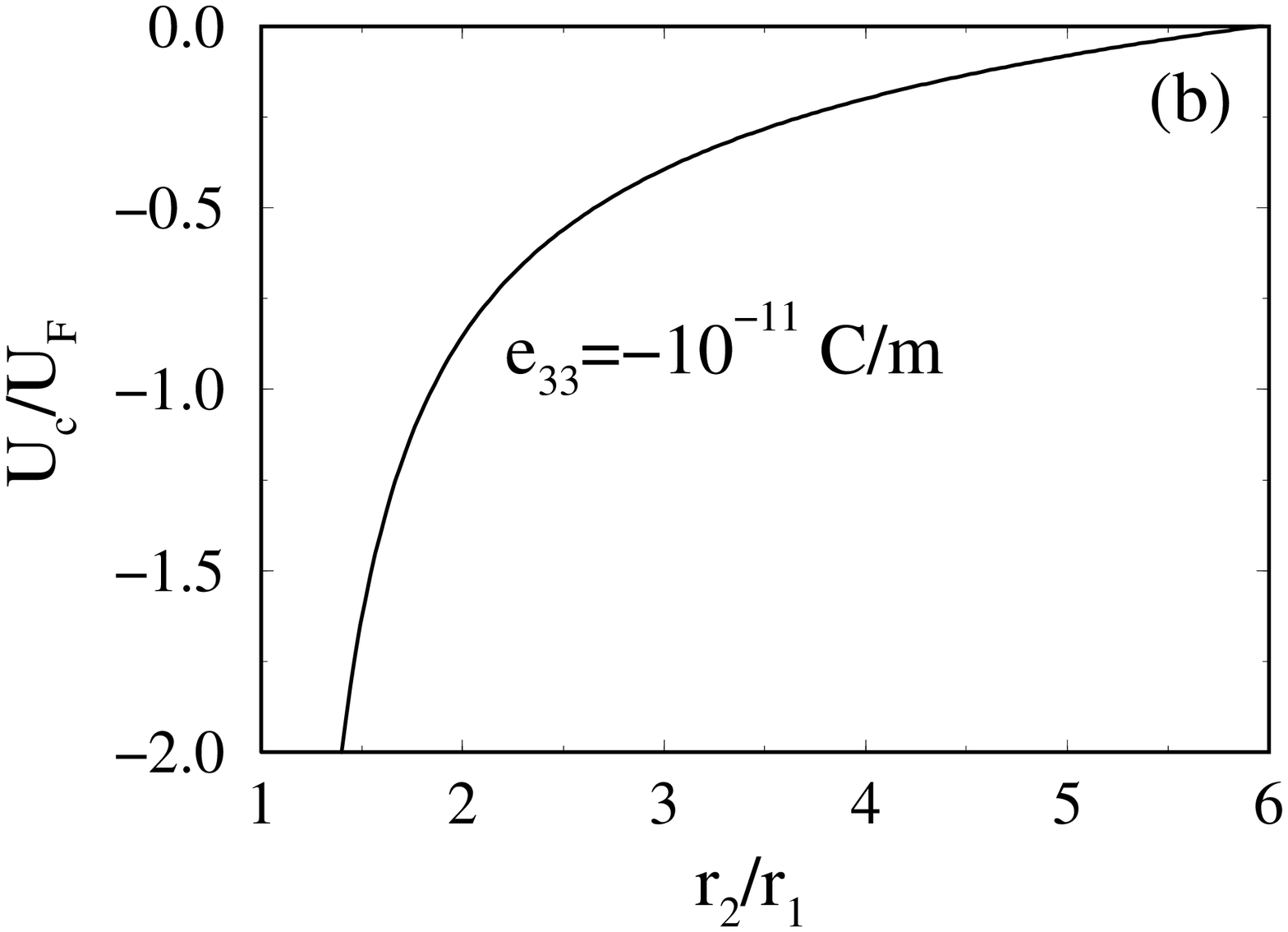}
\end{center}
\vspace*{-1.0cm}
\caption{Critical voltage versus ratio of the radii for MBBA with
homeotropic (a) and planar (b) anchoring.}
\label{fig2}
\end{figure}

For a nematic liquid crystal with negative dielectric anisotropy 
($\epsilon_a<0$) the flexoelectric effect reduces the critical
voltage for the orientational transition.
Using the MBBA material parameters at $25^\circ$C
$K_1=6.66\cdot 10^{-12}$~N,
$K_2=4.2\cdot 10^{-12}$~N,
$K_3=8.61\cdot 10^{-12}$~N,
$\epsilon_a=-0.53$
and taking for the flexocoefficients the order of magnitude
$e_{11}+e_{33}=e_{11}=-10^{-11}$~C/m \cite{DMD82,MD85}
we found that the transition corresponding to $\delta n_{z}$ 
perturbation has lower threshold [Fig.~\ref{fig2}(a)].
In this case above $r_2/r_1\approx 35.6$ the orientational transition 
occurs without electric field.

\vspace{12pt}
\noindent
{\bf II)} Planar anchoring with boundary conditions
\begin{equation}
{\bf \hat n}(r=r_1) = {\bf \hat n}(r=r_2) = (0,1,0) \;.
\end{equation} 
Similar to the case of homeotropic orientation, small deviations from the
exact solution ${\bf \hat n}_0=(0,1,0)$ of (\ref{torqbal}) 
can be written as
\begin{equation}
{\bf \hat n} = {\bf \hat n}_0 + \delta{\bf \hat n} = 
\left[ \delta n_{r}; 1-\frac12 (\delta n_{r}^2 + \delta n_{z}^2); \delta n_{z}  \right]
\end{equation} 
up to second order in the perturbations $\delta n_{r}$, $\delta n_{z}$ which depend 
only on $r$.
Then the change of total free energy up to the second order in
$\delta {\bf \hat n}$ is 
\begin{eqnarray}
&& \Delta {\cal F}_{plan} \equiv 
{\cal F}({\bf \hat n}_0) - {\cal F}({\bf \hat n}_0+\delta {\bf \hat n}) = 
\nonumber\\
&& = 2 \pi \int_{r_1}^{r_2} 
\Bigl\{ 
\frac12 K_1 [\delta n_{r}'^2 + \frac{1}{r^2}\delta n_{r}^2] +
\frac12 K_2 [\delta n_{z}'^2 + \frac{1}{r^2}\delta n_{z}^2] - 
\nonumber\\
&& \qquad \qquad \,\,\, \frac12 K_3 \frac{1}{r^2}(\delta n_{r}^2 + \delta n_{z}^2) - 
\frac12 \epsilon_0\epsilon_a {\tilde U}^2 \frac{1}{r^2} \delta n_{r}^2 - 
\nonumber\\
&& \qquad \qquad \,\,\, {\tilde U} e_{11} \frac{1}{r^2} \delta n_{r}^2 -  
{\tilde U} e_{33} \frac{1}{r^2} (\delta n_{r}^2 + \delta n_{z}^2)
\Bigr\} r d r \; ,
\end{eqnarray} 
where the boundary conditions $\delta {\bf \hat n}(r=r_1)=\delta {\bf \hat n}(r=r_2)=0$ 
were taken into account.

\begin{figure} 
\begin{center} 
\vspace*{-0.3cm} 
\hspace*{0cm} 
\epsfxsize=7cm 
\epsfbox{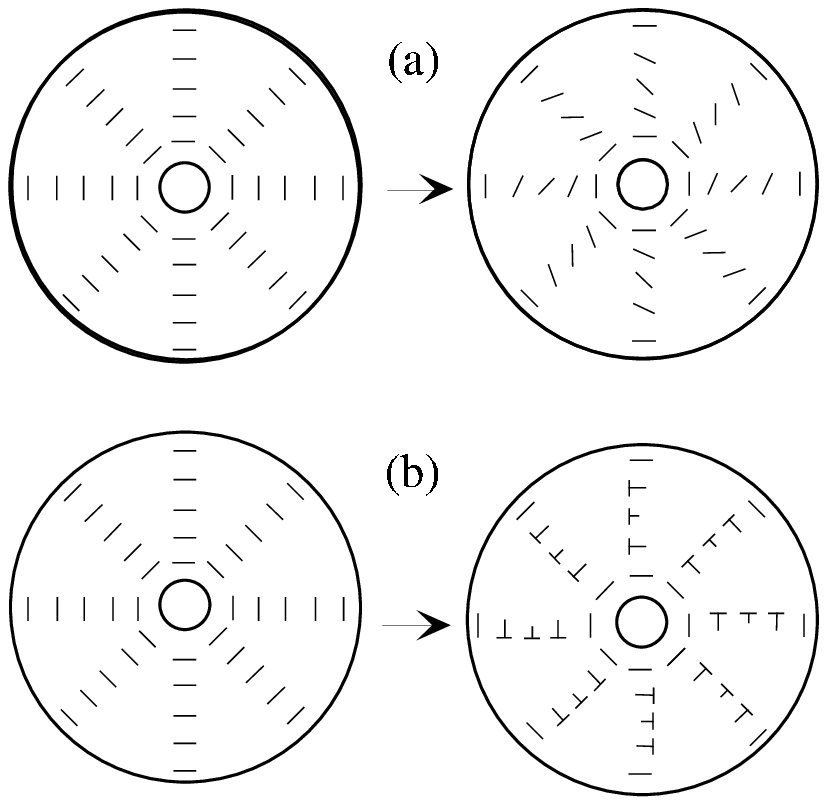} 
\end{center} 
\vspace*{-0.8cm}
\caption{Orientational transition for $\delta n_{r}$ (a) 
and $\delta n_{z}$ (b).} 
\label{plan} 
\end{figure} 

The instability conditions can be obtained from the Euler-Lagrange equations
\begin{eqnarray}
\label{dnr}
&& K_1 \frac{d^2}{d \alpha^2}\delta n_{r} +
[K_3 - K_1 + \epsilon_0\epsilon_a {\tilde U}^2 + 2{\tilde U}(e_{11}+e_{33})] \delta n_{r} = 0 \; , \;\; \\
\label{dnz_p}
&& K_2 \frac{d^2}{d \alpha^2}\delta n_{z} + 
[2K_3 - K_2 + 2{\tilde U}e_{33}] \delta n_{z} = 0 \; .
\end{eqnarray}
Here again the equations (\ref{dnr}), (\ref{dnz_p}) are uncoupled and one has two independent 
conditions for the instability of initial planar orientation with respect to $\delta n_{r}$ and
$\delta n_{z}$ perturbations (Fig.~\ref{plan}).
The critical voltage $U_c$ for the orientational transition can be found from the solution of 
(\ref{dnr}), (\ref{dnz_p})
\begin{eqnarray}
\label{Ucpdnr}
\delta n_{r}\;: &&
{\rm sign}(\epsilon_a)\left(\frac{U_c}{U_{F_1}}\right)^2 + 
\left(\frac{U_c}{U_{F_1}}\right) \frac{2(e_{11}+e_{33})\ln(r_2/r_1)}{\pi \sqrt{K_1\epsilon_0|\epsilon_a|}} = 
\nonumber\\
&& \qquad \;\;\;\;\,\, = 1 - \frac{K_3-K_1}{K_1}\left(\frac{\ln(r_2/r_1)}{\pi}\right)^2 \, ,
\\
\label{Ucpdnz}
\delta n_{z}\;: &&
 \left(\frac{U_c}{U_{F_2}}\right) \frac{2e_{33}\ln(r_2/r_1)}{\pi \sqrt{K_2\epsilon_0|\epsilon_a|}} = 
\nonumber\\
&&  \qquad \;\;\;\;\,\, = 1 - \frac{2K_3-K_2}{K_2}\left(\frac{\ln(r_2/r_1)}{\pi}\right)^2 \, ,
\end{eqnarray}
where $U_{F_1}=\pi\sqrt{K_1/\epsilon_0|\epsilon_a|}$, 
$U_{F_2}=\pi\sqrt{K_2/\epsilon_0|\epsilon_a|}$ are the threshold 
voltages for the splay and twist Fr\'eedericksz transition in the case of a plane 
NLC layer, respectively.
Without electric field one has the orientational transitions at the critical
radius ratio
$\ln(r_2/r_1)=\pi\sqrt{K_1/(K_3-K_1)}$ [Fig.~\ref{plan}(a)] and 
$\ln(r_2/r_1)=\pi\sqrt{K_2/(2K_3-K_2)}$ [Fig.~\ref{plan}(b)].

For NLC with $\epsilon_a>0$ the flexoelectric effect reduces the
critical voltage for the transition corresponding to $\delta n_r$
perturbations similar to that found in the case of homeotropically
oriented nematics with $\epsilon_a<0$ [Fig.~\ref{fig2}(a)].
Equation (\ref{Ucpdnz}) demonstrates a new feature of the cylindrical 
geometry, namely, the possibility
of an orientational transition caused purely by the flexoelectric effect.
In Figure~\ref{fig2}(b) the threshold voltage corresponding to the
flexoelectric instability ($\delta n_z$ perturbation) as a function of 
the ratio of the radii is plotted for the MBBA material parameters
and $e_{33}=-10^{-11}$~C/m.
Above $r_2/r_1\approx 5.96$ the transition takes place without
electric field.

\vspace{12pt}
Thus, the analysis of the orientational transitions in a NLC confined
between two coaxial cylinders shows the strong influence of the
flexoelectric effect on the critical voltage.
For planar boundary conditions a pure flexoelectric instability
is found.
In this case the polarity of critical voltage depends only on the 
sign of the flexo-coefficient $e_{33}$ which allows 
to determine the value and the sign of $e_{33}$ in a simple experiment.
Choosing the radius ratio of the cylinders close to the critical 
(above which the transition occurs without electric field)
one can make the threshold voltage for the flexo\-electric-induced orientational transition very small
and reduce the possible influence of the electric current in a not sufficiently clean NLC 
materials.

\vspace{12pt}
\noindent
{\bf Acknowledgments}

\noindent
We thank L. Kramer for fruitful discussions and a critical reading
of the manuscript.
A.K. wishes to acknowledge the hospitality of the University of Bayreuth.
Financial support from DFG Grant Kr-690/14-1 and INTAS Grant 96-498 is 
gratefully acknowledged.
M.Kh. is also grateful to the INTAS for fellowship grant YSF 99-4035.

\setlength{\baselineskip}{12pt}

\end{document}